# Metropolis Sampling

Luca Martino*, Victor Elvira$^\diamond$
* Universidad de Valencia (Spain).
$^\diamond$ Télécom Lille, CRIStAL laboratory (France).

**Abstract**

Monte Carlo (MC) sampling methods are widely applied in Bayesian inference, system simulation and optimization problems. The Markov Chain Monte Carlo (MCMC) algorithms are a well-known class of MC methods which generate a Markov chain with the desired invariant distribution. In this document, we focus on the Metropolis-Hastings (MH) sampler, which can be considered as the atom of the MCMC techniques, introducing the basic notions and different properties. We describe in details all the elements involved in the MH algorithm and the most relevant variants. Several improvements and recent extensions proposed in the literature are also briefly discussed, providing a quick but exhaustive overview of the current Metropolis-based sampling's world.

**Keywords:** Monte Carlo methods, Metropolis-Hastings algorithm, Markov Chain Monte Carlo (MCMC).

## 1 Problem Statement

In many practical applications, the goal consists in inferring a variable of interest, $\mathbf{x} = [x_1, \ldots, x_D] \in \mathbb{R}^D$, given a set of observations or measurements, $\mathbf{y} \in \mathbb{R}^P$. In the Bayesian framework [52], the total knowledge about the parameters, after the data have been observed, is represented by the posterior probability density function (pdf), i.e.,

$$\bar{\pi}(\mathbf{x}) = p(\mathbf{x}|\mathbf{y}) = \frac{\ell(\mathbf{y}|\mathbf{x})g(\mathbf{x})}{Z(\mathbf{y})}, \tag{1}$$

where $\ell(\mathbf{y}|\mathbf{x})$ denotes the likelihood function (i.e., the observation model), $g(\mathbf{x})$ is the prior probability density function (pdf) and $Z(\mathbf{y})$ is the marginal likelihood (a.k.a., Bayesian evidence) [10, 83]. In general $Z(\mathbf{y})$ is unknown, and it is possible to evaluate the unnormalized target function,

$$\pi(\mathbf{x}) = \ell(\mathbf{y}|\mathbf{x})g(\mathbf{x}). \tag{2}$$

The analytical study of the posterior density $\bar{\pi}(\mathbf{x}) \propto \pi(\mathbf{x})$ is often unfeasible and integrals involving $\bar{\pi}(\mathbf{x})$ are typically intractable [10, 14, 83]. For instance, one might be interested in the estimation

Table 1: Summary of the notation.

| | |
|---|---|
| $D$ | dimension of the inference problem, $\mathbf{x} \in \mathbb{R}^D$. |
| $T$ | total number of generated samples. |
| $\mathbf{x}$ | variable of interest; parameter to be inferred, $\mathbf{x} = [x_1, \ldots, x_D]$. |
| $\mathbf{y}$ | observed data. |
| $\ell(\mathbf{y}\|\mathbf{x})$ | likelihood function. |
| $g(\mathbf{x})$ | prior density. |
| $\bar{\pi}(\mathbf{x})$ | posterior (target) density $\bar{\pi}(\mathbf{x}) = p(\mathbf{x}\|\mathbf{y})$. |
| $\pi(\mathbf{x})$ | posterior (target) function, $\pi(\mathbf{x}) = \ell(\mathbf{y}\|\mathbf{x})g(\mathbf{x}) \propto \bar{\pi}(\mathbf{x})$. |
| $q(\mathbf{x})$ | proposal density. |
| $I$ | integral to be approximated, $I = E_\pi[f(\mathbf{x})]$. |

of

$$I = E_\pi[f(\mathbf{x})] = \int_{\mathbb{R}^D} f(\mathbf{x})\bar{\pi}(\mathbf{x})d\mathbf{x}, \qquad (3)$$

where $f(\mathbf{x})$ is a squared integrable function w.r.t. $\bar{\pi}$, i.e. $E_\pi[f(\mathbf{x})^2] < \infty$. Table 1 summarizes the notation.

## 1.1 Monte Carlo integration

In many practical scenarios, the integral $I$ cannot be computed in a closed form, and numerical approximations are typically required. Many deterministic quadrature methods are available in the literature [1, 4, 18, 51, 80]. However, as the dimension $D$ of the inference problem grows ($\mathbf{x} \in \mathbb{R}^D$), the deterministic quadrature schemes become less efficient. In this case, a common approach consists of approximating the integral $I$ in Eq. (3) by using Monte Carlo (MC) quadrature [25, 47, 33, 50, 57, 70, 85]. Namely, considering $T$ are independent and identically distributed (i.i.d.) samples drawn from the posterior target pdf, i.e. $\mathbf{x}^{(1)}, \ldots, \mathbf{x}^{(T)} \sim \bar{\pi}(\mathbf{x})$,[1] we can build the consistent estimator

$$\widehat{I}_T = \frac{1}{T}\sum_{t=1}^T f(\mathbf{x}^{(t)}) \xrightarrow{p} I, \qquad (4)$$

i.e., $\widehat{I}_T$ converges in probability to $I$ due to the weak law of large numbers. In other words, for any positive number $\epsilon > 0$, we have $\lim_{T \to \infty} \Pr(|\widehat{I}_T - I| > \epsilon) = 0$. Note also that $E_\pi[\widehat{I}_T] = I$, i.e. the estimator is unbiased. The approximation $\widehat{I}_T$ is known as a direct (or ideal) Monte Carlo estimator if the samples $\mathbf{x}^{(t)}$ are i.i.d. from $\bar{\pi}$.

---

[1] In this work, for simplicity, we use the same notation for denoting a random variable or one realization of a random variable.

### 1.1.1 Efficiency of the direct Monte Carlo estimator

Since the samples are i.i.d., the variance of the estimator $\widehat{I}_T$ is given by

$$\mathrm{Var}\left[\widehat{I}_T\right] = \frac{1}{T^2} \sum_{t=1}^{T} \mathrm{Var}\left[f(\mathbf{x}^{(t)})\right], \tag{5}$$

$$= \frac{\sigma_f^2}{T}, \tag{6}$$

where $\sigma_f^2 = \mathrm{Var}\left[f(\mathbf{x})\right]$, and i.i.d. $\mathbf{x}^{(1)}, \ldots, \mathbf{x}^{(T)} \sim \bar{\pi}(\mathbf{x})$. Therefore, the variance of $\widehat{I}_T$ depends on the functions $f$ and $\bar{\pi}$ (i.e., the variance of the random variable $\mathbf{z} = f(\mathbf{x})$ with $\mathbf{x} \sim \bar{\pi}(\mathbf{x})$) and decreases linearly with the number $T$ of i.i.d. samples. Due to the central limit theorem, the distribution of $\widehat{I}_T$ is approximated by $\mathcal{N}\left(I, \frac{\sigma_f^2}{T}\right)$ when $T \to \infty$.

## 1.2 Markov chain Monte Carlo (MCMC) methods

Unfortunately, drawing independent samples from $\bar{\pi}(\mathbf{x})$ is in general not possible, and alternative approaches, e.g., Markov chain Monte Carlo (MCMC) algorithms, are needed [24, 37, 55, 85]. An MCMC method generates an ergodic Markov chain with invariant (a.k.a., stationary) density given by the posterior pdf $\bar{\pi}(\mathbf{x})$ [53]. Specifically, given a starting state $\mathbf{x}^{(0)}$, a sequence of *correlated* samples is generated, $\mathbf{x}^{(0)} \to \mathbf{x}^{(1)} \to \mathbf{x}^{(2)} \to \ldots \to \mathbf{x}^{(T)}$. Even if the samples are now correlated, the estimator

$$\widetilde{I}_T = \frac{1}{T} \sum_{t=1}^{T} f(\mathbf{x}^{(t)}) \tag{7}$$

is consistent, regardless the starting vector $\mathbf{x}^{(0)}$ [91].[2] With respect to the direct Monte Carlo approach using i.i.d. samples, the application of an MCMC algorithm entails a loss of efficiency of the estimator $\widehat{I}_T$, if the samples are positively correlated. In other words, to achieve a given variance obtained with the direct Monte Carlo estimator, it is necessary to generate more samples. In order to clarify this consideration, let us define the auto-covariance at lag $k$

$$\gamma_k = \mathrm{cov}\left[f(\mathbf{x}^{(t)}), f(\mathbf{x}^{(t+k)})\right],$$

so that $\gamma_0 = \sigma_f^2 = \mathrm{Var}\left[f(\mathbf{x})\right]$. Then, the autocorrelation at lag $k$ is defined as

$$\rho_k = \frac{\mathrm{cov}\left[f(\mathbf{x}^{(t)}), f(\mathbf{x}^{(t+k)})\right]}{\mathrm{Var}\left[f(\mathbf{x})\right]},$$

$$= \frac{\gamma_k}{\sigma_f^2}. \tag{8}$$

Note that $\rho_0 = 1$ and $\rho_0 \geq |\rho_k|$ for all $k > 0$ ($k \in \mathbb{N}$). It can be shown that the variance of the estimator $\widetilde{I}_T$ in Eq. (7) when the samples have autocorrelation function $\rho_k$ is [29]

$$\mathrm{Var}\left[\widetilde{I}_T\right] = \frac{\sigma_f^2}{T}\left(1 + \frac{2}{T}\sum_{k=1}^{T-1}(T-k)\rho_k\right). \tag{9}$$

---
[2]Recall we are assuming that the Markov chain is ergodic and hence the starting value is forgotten.

Moreover, it is possible to write that

$$\lim_{T \to \infty} T \text{Var}\left[\widetilde{I}_T\right] = \sigma_f^2 \left(1 + 2\sum_{k=1}^{\infty} \rho_k\right), \tag{10}$$

if the series of $\rho_k$ is convergent. In order to quantify the loss of efficiency w.r.t. the direct Monte Carlo approach (see Eq. (6)), we can define the *Effective Sample Size* (ESS) for an MCMC technique

$$T_{eff} = \lim_{T \to \infty} T \frac{\text{Var}\left[\widehat{I}_T\right]}{\text{Var}\left[\widetilde{I}_T\right]}, \tag{11}$$

$$= \frac{T}{1 + 2\sum_{k=1}^{\infty} \rho_k}, \tag{12}$$

which can be thought of the number of independent samples needed to obtain a variance of $\text{Var}\left[\widetilde{I}_T\right]$, i.e., $\text{Var}\left[\widehat{I}_{T_{eff}}\right] = \text{Var}\left[\widetilde{I}_T\right]$. Since, in general, the MCMC algorithm induces positive correlation among the samples, then $T_{eff} \leq T$ [29]. Note that, the equality $T_{eff} = T$ is achieved if there is zero correlation among the samples. Thus, considering the same computational cost, an MCMC technique provides better performance than other MCMC method if the generated samples present less correlation.

**Burn-in.** Another consequence of the correlation is the burn-in period that the chain requires before converging to the invariant distribution. Therefore a certain number of initial samples should be discarded, i.e., not included in the resulting estimator [7, 85]. However, the length of the burn-in period is in general unknown. Several studies in order to estimate the length of the burn-in period [16, 32] or to avoid it [81] can be found in the literature.

**Example 1.** *Let us consider the study of an invariant probability mass function (pmf) of a Markov chain in a discrete space. In this case, we consider $\mathbf{x} \in \{1, 2, 3\}$ and a $3 \times 3$ transition matrix (a.k.a., as a kernel matrix) of the chain,*

$$\mathbf{K} = \{k_{i,j}\} = \begin{bmatrix} 0.3 & 0.3 & 0 \\ 0.7 & 0.1 & 0.5 \\ 0 & 0.6 & 0.5 \end{bmatrix}. \tag{13}$$

*where $k_{i,j}$ represents the probability of the transition from i-th to the j-th state. For instance, $k_{1,2} = 0.7$ is the probability of jumping from state 1 to state 2. Let us consider the eigenvalue problem*

$$\mathbf{Kp} = \lambda \mathbf{p},$$

*where $\lambda \in \mathbb{R}$ and $\mathbf{p}$ is $3 \times 1$ generic vector. The eigenvalues of $\mathbf{K}$ are $\lambda_1 = 1$, $\lambda_2 = -0.4772$ and $\lambda_3 = 0.3772$. The eigenvector associated to $\lambda_1 = 1$ is the invariant probability mass function (pmf)*

$$\bar{\boldsymbol{\pi}} = [0.1630, \ 0.3804, \ 0.4565]^\top.$$

*This pmf is called invariant since*

$$\mathbf{K}\bar{\boldsymbol{\pi}} = \bar{\boldsymbol{\pi}},$$

*i.e., for a dynamical point of view, $\bar{\boldsymbol{\pi}}$ is a fixed point for the application $\mathbf{K}$. Moreover, the fixed point $\bar{\boldsymbol{\pi}}$ is attractive and stable. Indeed, let us consider now the starting pmf $\mathbf{p}_0 = [0, \ 0, \ 1]^\top$, and the following recursion*

$$\mathbf{p}_{t+1} = \mathbf{K}\mathbf{p}_t, \quad \text{for} \quad t = 0, 1, 2, \ldots,$$

*that can be also written as $\mathbf{p}_{t+1} = \mathbf{K}^t \mathbf{p}_0$. We can observe that*

$$\begin{aligned}
\mathbf{p}_1 &= \mathbf{K}\mathbf{p}_0 = [0, \ 0.5, \ 0.5]^\top, \\
\mathbf{p}_2 &= \mathbf{K}\mathbf{p}_1 = [0.50, \ 0.30, \ 0.55]^\top, \\
&\vdots \\
\mathbf{p}_{13} &= \mathbf{K}\mathbf{p}_{12} = [0.1630, \ 0.3805, \ 0.4565]^\top, \\
\mathbf{p}_{14} &= \mathbf{K}\mathbf{p}_{13} = [0.1630, \ 0.3804, \ 0.4565]^\top \approx \bar{\boldsymbol{\pi}}, \\
&\vdots \\
\bar{\boldsymbol{\pi}} &= \mathbf{K}\bar{\boldsymbol{\pi}},
\end{aligned}$$

*i.e., after 14 steps (i.e., the burn-in period), the chain has virtually reached the invariant pmf $\bar{\boldsymbol{\pi}}$. It is possible to show that*

$$\lim_{t \to \infty} \mathbf{p}_t = \bar{\boldsymbol{\pi}}, \quad \text{and} \quad \lim_{t \to \infty} \mathbf{K}^t = [\bar{\boldsymbol{\pi}}, \ \bar{\boldsymbol{\pi}}, \ \bar{\boldsymbol{\pi}}]$$

*i.e., the matrix $\mathbf{K}^t$ converges, as $t$ grows, to a matrix whose columns are exactly $\bar{\boldsymbol{\pi}}$, so that $\bar{\boldsymbol{\pi}} = \mathbf{K}^\infty \mathbf{p}_0$ for any possible starting pmf $\mathbf{p}_0$.*

## 2  Metropolis-Hastings (MH) algorithm

**Algorithm.** One of the most popular and widely applied MCMC algorithm is the Metropolis-Hastings (MH) method [42, 71, 29, 55, 84]. Nicolas Metropolis (physicist and mathematician) came to Los Alamos (New Mexico) in 1943, during the World War II, joining a group of scientists working on mathematical physics and the atomic bomb. In 1953, they published a first version of the algorithm in the *Journal of Chemical Physics* [71]. The algorithm was generalized by W. Hastings in 1970 [42]. In order to apply the MH method, the only requirement is to be able to evaluate point-wise a function proportional to the target, i.e., $\pi(\mathbf{x}) \propto \bar{\pi}(\mathbf{x})$. For the sake of simplicity, we assume that $\pi(\mathbf{x}) > 0$ for all $\mathbf{x} \in \mathbb{R}^D$. A proposal density (a pdf which is easy to draw from) is denoted as $q(\mathbf{x}|\mathbf{x}^{(t-1)}) > 0$, with $\mathbf{x}, \mathbf{x}^{(t-1)} \in \mathbb{R}^D$. Below, we describe the standard MH algorithm in detail.

> **The Metropolis-Hastings algorithm**
>
> 1. Choose an initial state $\mathbf{x}^{(0)}$.
>
> 2. For $t = 1, \ldots, T$:
>
>    (a) Draw a sample $\mathbf{z}' \sim q(\mathbf{x}|\mathbf{x}^{(t-1)})$.
>
>    (b) Accept the new state, $\mathbf{x}^{(t)} = \mathbf{z}'$, with probability
>
>    $$\alpha(\mathbf{x}^{(t-1)}, \mathbf{z}') = \min\left[1, \frac{\pi(\mathbf{z}')q(\mathbf{x}^{(t-1)}|\mathbf{z}')}{\pi(\mathbf{x}^{(t-1)})q(\mathbf{z}'|\mathbf{x}^{(t-1)})}\right]. \tag{14}$$
>
>    Otherwise, set $\mathbf{x}^{(t)} = \mathbf{x}^{(t-1)}$.

The algorithm returns the sequence of states $\{\mathbf{x}^{(1)}, \mathbf{x}^{(2)}, \ldots, \mathbf{x}^{(t)}, \ldots, \mathbf{x}^{(T)}\}$ or a subset of them removing the burn-in period if an estimation of its length is available. We can see that the next state $\mathbf{x}^{(t)}$ can be the proposed sample $\mathbf{z}'$ (with probability $\alpha$) or the previous state $\mathbf{x}^{(t-1)}$ (with probability $1 - \alpha$). Under some mild regularity conditions, when $t$ grows, the pdf of the current state $\mathbf{x}^{(t)}$ converges to the target density $\bar{\pi}(\mathbf{x})$ [85]. The MH algorithm satisfies the so-called detailed balance condition which is sufficient to guarantee that the output chain is ergodic and has $\bar{\pi}$ as stationary distribution [29, 55, 85] (see proof in Section 2.3).

## 2.1 Elements in the MH method

Observe the MH algorithm involves the following three elements:

- the proposal density $q(\mathbf{z}|\mathbf{x})$,
- the acceptance probability $\alpha(\mathbf{x}, \mathbf{z})$,
- and the target function $\pi(\mathbf{x})$.

All of them can be varied, improved or extended in order to improve the performance of the algorithm, and always ensuring the ergodicity of the chain (see Sections 2.2, 2.4 and 3). The proposal pdf $q$ should be chosen as close as possible to the target $\pi$. Different acceptance probabilities $\alpha$ could be used, and in generalized MH methods suitable acceptance functions must be designed in order to guarantee the ergodicity of the generated chain. Furthermore, in some specific scenario, an artificial increase of the parameter space can be useful for improving the performance. More specifically, higher dimensional target functions $\pi_g$ are studied in some cases. The extended target $\pi_g$ is built in order to have the true posterior pdf $\pi$ as a marginal density. Other important considerations about the MH method are:

- the inference is directly performed in the $D$-dimensional space, i.e., using a *block* approach (unlike in a component-wise strategy; see Section 3),

- only one sample $\mathbf{z}'$ is proposed as new possible state,

- the dimensionality of the problem $D$ is considered fixed,
- the only requirement is to be able to evaluate $\pi(\mathbf{x})$.

In the rest of the work, we broaden the discussion about the flexibility of the MH algorithm and its extensions.

## 2.2 Particular choices of the proposal form

Some interesting cases, depending on the choice of the proposal density, are discussed below:

1. **Symmetric proposal:** if the proposal satisfies the equality $q(\mathbf{z}|\mathbf{x}) = q(\mathbf{x}|\mathbf{z})$, then the acceptance probability is simplified, i.e.,

$$\alpha(\mathbf{x}, \mathbf{z}) = \min\left[1, \frac{\pi(\mathbf{z})}{\pi(\mathbf{x})}\right], \quad (15)$$

depending only on the evaluation of the target function $\pi$. This special case is interesting since it clarifies the strict relationship between the sampling and optimization problems. Indeed, in this case, we have

$$\alpha(\mathbf{x}, \mathbf{z}) = \begin{cases} 1, & \text{if } \pi(\mathbf{z}) \geq \pi(\mathbf{x}), \\ \frac{\pi(\mathbf{z})}{\pi(\mathbf{x})}, & \text{if } \pi(\mathbf{z}) < \pi(\mathbf{x}), \end{cases} \quad (16)$$

i.e., the samples providing higher values of $\pi$ are always accepted, whereas the samples providing smaller values of $\pi$ are only accepted according to a certain probability. Figure 1 depicts this scenario.

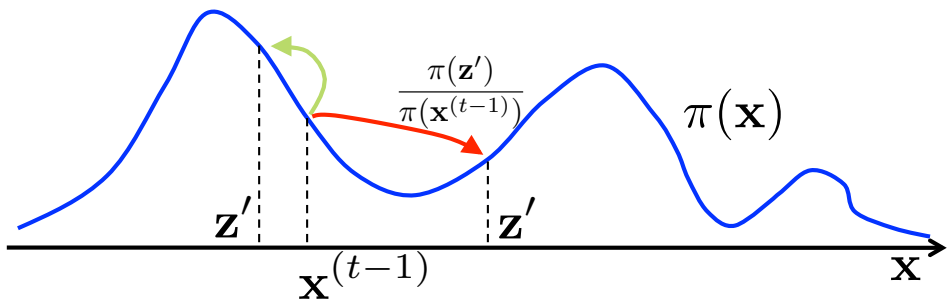

Figure 1: With a symmetric proposal, the "uphill" samples $\mathbf{z}'$ are always accepted as next state. The "downhill" jumps are accepted with probability $\alpha = \frac{\pi(\mathbf{z}')}{\pi(\mathbf{x}^{(t-1)})}$.

**Variant for Optimization.** The *simulated annealing* method is an optimization algorithm [49, 59, 44] which can be considered a variant of the MH method where the target $\pi$ is *tempered* (i.e., scaled) in order to decrease the probability of accepting the new points $\mathbf{z}$'s with smaller values $\pi(\mathbf{z})$ w.r.t. the current state $\mathbf{x}$. More specifically, in simulated annealing, the $\alpha$ is modified as

$$\alpha(\mathbf{x}, \mathbf{z}) = \min\left[1, \left(\frac{\pi(\mathbf{z})}{\pi(\mathbf{x})}\right)^{\gamma_t}\right], \qquad (17)$$

where $\gamma_t \in [1, \infty)$ and for optimization purposes is an increasing function of $t$, i.e.,

$$\gamma_0 = 1 \leq \gamma_1 \leq \gamma_2 \leq \cdots \leq \gamma_t < \infty.$$

Indeed, for $\gamma_t > 1$ and $\mathbf{z} \neq \mathbf{x}$, if $\pi(\mathbf{z}) > \pi(\mathbf{x})$ we have $\left(\frac{\pi(\mathbf{z})}{\pi(\mathbf{x})}\right)^{\gamma_t} > \frac{\pi(\mathbf{z})}{\pi(\mathbf{x})} > 1$ whereas, if $\pi(\mathbf{z}) < \pi(\mathbf{x})$, we have $\left(\frac{\pi(\mathbf{z})}{\pi(\mathbf{x})}\right)^{\gamma_t} < \frac{\pi(\mathbf{z})}{\pi(\mathbf{x})}$ so that the ratio $\left(\frac{\pi(\mathbf{z})}{\pi(\mathbf{x})}\right)^{\gamma_t}$ is closer to zero.

2. **Independent proposal:** if the proposal pdf is independent on the previous state of the chain, i.e., $q(\mathbf{z}|\mathbf{x}) = q(\mathbf{x})$, the acceptance probability is

$$\alpha(\mathbf{x}, \mathbf{z}) = \min\left[1, \frac{\pi(\mathbf{z})q(\mathbf{x})}{\pi(\mathbf{x})q(\mathbf{z})}\right], \qquad (18)$$

$$= \min\left[1, \frac{w(\mathbf{z})}{w(\mathbf{x})}\right], \qquad (19)$$

where $w(\mathbf{x}) = \frac{\pi(\mathbf{x})}{q(\mathbf{x})}$ is the weight associated to the sample $\mathbf{x}$ in the standard importance sampling technique [17, 27, 28, 26, 64, 63].

3. **Optimal proposal:** if $q(\mathbf{x}) = \bar{\pi}(\mathbf{x})$, i.e., the proposal coincides with the target pdf (recall $\bar{\pi}(\mathbf{x}) \propto \pi(\mathbf{x})$), we have

$$\alpha(\mathbf{x}, \mathbf{z}) = 1, \qquad (20)$$

hence all the proposed samples are accepted since they are distributed as $\bar{\pi}(\mathbf{x})$. The produced states are independent and, as consequence, they have zero correlation among each other.

4. **Random walk proposal:** this is the case when the proposal pdf can be expressed as $q(\mathbf{z}|\mathbf{x}) = q(\mathbf{z} - \mathbf{x})$, i.e., $\mathbf{x}$ plays the role of a location parameter (as the mean). This choice presents an interesting explorative behavior since the location of the proposal follows a random walk around the parameter space, according to the states of chain.

**Example 2.** *An example of a both symmetric and random walk proposal is*

$$q(\mathbf{z}|\mathbf{x}) \propto \exp\left(-\frac{(\mathbf{z} - \mathbf{x})^2}{2\sigma^2}\right), \qquad (21)$$

*i.e., a Gaussian density with mean $\mathbf{x}$ and covariance matrix $\sigma^2 \mathbf{I}_D$.*[3]

---
[3] With $\mathbf{I}_D$, we have denoted the identity matrix of dimensions $D \times D$.

## 2.3 Invariant distribution of the MH algorithm

Let us denote as $K(\mathbf{z}|\mathbf{x})$ the transition function (or *kernel*) that determines the probability of moving from the state $\mathbf{x}$ to the state $\mathbf{z}$ of an MCMC algorithm. A generic MCMC technique has $\bar{\pi}(\mathbf{x})$ as an invariant (or stationary) distribution if its kernel satisfies

$$\int_{\mathbb{R}^D} K(\mathbf{z}|\mathbf{x})\bar{\pi}(\mathbf{x})d\mathbf{x} = \bar{\pi}(\mathbf{z}), \tag{22}$$

i.e., $\bar{\pi}(\mathbf{x})$ plays the role of an eigenfunction w.r.t. the integral operator including $K(\mathbf{z}|\mathbf{x})$ associated to an eigenvalue 1 (see Ex. 1 for the discrete case). A sufficient condition which implies the equation above is the detailed balance condition [85],

$$\bar{\pi}(\mathbf{x})K(\mathbf{z}|\mathbf{x}) = \bar{\pi}(\mathbf{z})K(\mathbf{x}|\mathbf{z}). \tag{23}$$

If the condition above if fulfilled, $\bar{\pi}$ is invariant w.r.t. $K$ and the chain is also reversible [29, 85]. This condition is equivalent to Eq. (22) since by integrating both sides of (23) w.r.t. $\mathbf{x}$, we obtain

$$\int_{\mathbb{R}^D} \bar{\pi}(\mathbf{x})K(\mathbf{z}|\mathbf{x})d\mathbf{x} = \int_{\mathbb{R}^D} \bar{\pi}(\mathbf{z})K(\mathbf{x}|\mathbf{z})d\mathbf{x}, \tag{24}$$

$$\int_{\mathbb{R}^D} \bar{\pi}(\mathbf{x})K(\mathbf{z}|\mathbf{x})d\mathbf{x} = \bar{\pi}(\mathbf{z}). \tag{25}$$

In the following, we show that the MH technique yields a reversible chain, with invariant pdf $\bar{\pi}(\mathbf{x}) \propto \pi(\mathbf{x})$, by proving that it fulfills the detailed balance condition. The kernel of the MH method is

$$K(\mathbf{z}|\mathbf{x}) = \begin{cases} q(\mathbf{z}|\mathbf{x})\alpha(\mathbf{x},\mathbf{z}), & \text{for} \quad \mathbf{x} \neq \mathbf{z}, \\ \beta_r(\mathbf{x}), & \text{for} \quad \mathbf{x} = \mathbf{z}, \end{cases} \tag{26}$$

where

$$\beta_r(\mathbf{x}) = 1 - \int_{\mathbb{R}^D} q(\mathbf{v}|\mathbf{x})\alpha(\mathbf{x},\mathbf{v})d\mathbf{v},$$

is the rejection probability of a proposed generic state given that the previous state is $\mathbf{x}$ [5]. In the case $\mathbf{z} = \mathbf{x}$, the kernel is a Dirac delta function. For $\mathbf{z} \neq \mathbf{x}$, the balance condition is

$$\begin{aligned} \pi(\mathbf{x})K(\mathbf{z}|\mathbf{x}) &= \pi(\mathbf{x})q(\mathbf{z}|\mathbf{x})\alpha(\mathbf{x},\mathbf{z}), \\ &= \pi(\mathbf{x})q(\mathbf{z}|\mathbf{x})\min\left[1, \frac{\pi(\mathbf{z})q(\mathbf{x}|\mathbf{z})}{\pi(\mathbf{x})q(\mathbf{z}|\mathbf{x})}\right], \\ &= \min\left[\pi(\mathbf{x})q(\mathbf{z}|\mathbf{x}), \pi(\mathbf{z})q(\mathbf{x}|\mathbf{z})\right], \end{aligned} \tag{27}$$

where we have replaced the expression of $\alpha(\mathbf{x},\mathbf{z})$ in Eq. (14). We can observe that (27) is symmetric w.r.t. the variables $\mathbf{x}$ and $\mathbf{z}$ (i.e., they can be interchanged without varying the expression), then we can write $\pi(\mathbf{x})K(\mathbf{z}|\mathbf{x}) = \pi(\mathbf{z})K(\mathbf{x}|\mathbf{z})$, which is precisely the detailed balance condition. About the possible rates of convergence of an ergodic chain, see [78, 93].

## 2.4 Alternative acceptance functions

**General form.** There exist other alternative acceptance functions $\alpha(\mathbf{x}, \mathbf{z})$ which yield reversible MH kernels (i.e., that satisfy the balance condition). Let us define

$$r(\mathbf{x}, \mathbf{z}) = \frac{\pi(\mathbf{z})q(\mathbf{x}|\mathbf{z})}{\pi(\mathbf{x})q(\mathbf{z}|\mathbf{x})}, \tag{28}$$

and a function $h(r) : \mathbb{R}^+ \to [0, 1]$ that satisfies the following property

$$h(r) = rh(r^{-1}). \tag{29}$$

Then any acceptance probability defined as

$$\alpha(\mathbf{x}, \mathbf{z}) = h \circ r(\mathbf{x}, \mathbf{z}) = h(r(\mathbf{x}, \mathbf{z})), \tag{30}$$

is suitable in order to fulfill the balance condition (see [13] for a similar discussion). A more general formulation is

$$\alpha(\mathbf{x}, \mathbf{z}) = \lambda(\mathbf{x}, \mathbf{z})h(r(\mathbf{x}, \mathbf{z})), \tag{31}$$

where $\lambda(\mathbf{x}, \mathbf{z}) = \lambda(\mathbf{z}, \mathbf{x})$ is a symmetric function satisfying $0 \leq \alpha(\mathbf{x}, \mathbf{z}) \leq 1$ for all possible $\mathbf{x}$ and $\mathbf{z}$. Noting that $r(\mathbf{z}, \mathbf{x}) = \frac{1}{r(\mathbf{x},\mathbf{z})}$ and recalling Eq. (29), for the $\alpha$ functions defined in Eq. (30) we can also write

$$\begin{aligned} \alpha(\mathbf{x}, \mathbf{z}) = h(r(\mathbf{x}, \mathbf{z})) &= r(\mathbf{x}, \mathbf{z})h(r(\mathbf{x}, \mathbf{z})^{-1}), \\ &= r(\mathbf{x}, \mathbf{z})h(r(\mathbf{z}, \mathbf{x})), \end{aligned}$$

and finally we obtain

$$\alpha(\mathbf{x}, \mathbf{z}) = r(\mathbf{x}, \mathbf{z})\alpha(\mathbf{z}, \mathbf{x}). \tag{32}$$

Therefore, employing the property above, we can easily prove the detailed balance condition as follows

$$\begin{aligned} \pi(\mathbf{x})q(\mathbf{z}|\mathbf{x})\alpha(\mathbf{x}, \mathbf{z}) &= \pi(\mathbf{x})q(\mathbf{z}|\mathbf{x})r(\mathbf{x}, \mathbf{z})\alpha(\mathbf{z}, \mathbf{x}), \\ &= \pi(\mathbf{x})q(\mathbf{z}|\mathbf{x})\frac{\pi(\mathbf{z})q(\mathbf{x}|\mathbf{z})}{\pi(\mathbf{x})q(\mathbf{z}|\mathbf{x})}\alpha(\mathbf{z}, \mathbf{x}), \\ &= \pi(\mathbf{z})q(\mathbf{x}|\mathbf{z})\alpha(\mathbf{z}, \mathbf{x}). \end{aligned}$$

**Examples.** Considering Eq. (30) and choosing $h(r) = \min[1, z]$ we have again $\alpha(\mathbf{x}, \mathbf{z}) = \min\left[1, \frac{\pi(\mathbf{z})q(\mathbf{x}|\mathbf{z})}{\pi(\mathbf{x})q(\mathbf{z}|\mathbf{x})}\right]$. Choosing $h(r) = \frac{1}{1+\frac{1}{r}}$, considering Eq. (31) and recalling $r(\mathbf{x}, \mathbf{z}) = \frac{1}{r(\mathbf{z},\mathbf{x})}$, we obtain the so-called *Hastings' generalization* [42],

$$\alpha(\mathbf{x}, \mathbf{z}) = \frac{\lambda(\mathbf{x}, \mathbf{z})}{1 + \frac{\pi(\mathbf{x})q(\mathbf{z}|\mathbf{x})}{\pi(\mathbf{z})q(\mathbf{x}|\mathbf{z})}} = \frac{\lambda(\mathbf{x}, \mathbf{z})}{1 + \frac{1}{r(\mathbf{x},\mathbf{z})}} \tag{33}$$

$$= \frac{\lambda(\mathbf{x}, \mathbf{z})}{1 + r(\mathbf{z}, \mathbf{x})}, \tag{34}$$

$$= \frac{r(\mathbf{x}, \mathbf{z})\lambda(\mathbf{x}, \mathbf{z})}{r(\mathbf{x}, \mathbf{z}) + 1}, \tag{35}$$

With $\lambda(\mathbf{x},\mathbf{z}) = 1$, we obtain Barker's acceptance function [6],

$$\alpha(\mathbf{x},\mathbf{z}) = \frac{\pi(\mathbf{z})q(\mathbf{x}|\mathbf{z})}{\pi(\mathbf{z})q(\mathbf{x}|\mathbf{z}) + \pi(\mathbf{x})q(\mathbf{z}|\mathbf{x})}. \tag{36}$$

$$= \frac{\frac{\pi(\mathbf{z})}{q(\mathbf{z}|\mathbf{x})}}{\frac{\pi(\mathbf{z})}{q(\mathbf{z}|\mathbf{x})} + \frac{\pi(\mathbf{x})}{q(\mathbf{x}|\mathbf{z})}} = \frac{w(\mathbf{z}|\mathbf{x})}{w(\mathbf{z}|\mathbf{x}) + w(\mathbf{x}|\mathbf{z})}, \tag{37}$$

where, in the second line, we have multiplied numerator and denominator by $\frac{1}{q(\mathbf{x}|\mathbf{z})q(\mathbf{z}|\mathbf{x})}$ and we have denoted $w(\mathbf{x}|\mathbf{z}) = \frac{\pi(\mathbf{x})}{q(\mathbf{x}|\mathbf{z})}$ (in the same fashion of an importance sampling weight [57, 85]). Different discussions about the relationship of the Barker's acceptance function with resampling procedures and extensions with different numbers of proposed samples can be found in [75, 20, 65]. The Hastings' generalization in Eq. (33) includes also the standard acceptance function in Eq. (14) for $\lambda(\mathbf{x},\mathbf{z}) = 1 + \min[r(\mathbf{x},\mathbf{z}), r(\mathbf{z},\mathbf{x})]$. Indeed, replacing this choice of $\lambda(\mathbf{x},\mathbf{z})$ into Eq. (35), we have

$$\alpha(\mathbf{x},\mathbf{z}) = \frac{r(\mathbf{x},\mathbf{z})\lambda(\mathbf{x},\mathbf{z})}{r(\mathbf{x},\mathbf{z}) + 1},$$
$$= \frac{r(\mathbf{x},\mathbf{z})(1 + r(\mathbf{x},\mathbf{z}))}{r(\mathbf{x},\mathbf{z}) + 1} = r(\mathbf{x},\mathbf{z}), \quad \text{if} \quad r(\mathbf{x},\mathbf{z}) < r(\mathbf{z},\mathbf{x}),$$

and

$$\alpha(\mathbf{x},\mathbf{z}) = \frac{r(\mathbf{x},\mathbf{z})\lambda(\mathbf{x},\mathbf{z})}{r(\mathbf{x},\mathbf{z}) + 1},$$
$$= \frac{r(\mathbf{x},\mathbf{z})(1 + r(\mathbf{z},\mathbf{x}))}{r(\mathbf{x},\mathbf{z}) + 1},$$
$$= \frac{r(\mathbf{x},\mathbf{z})(1 + \frac{1}{r(\mathbf{x},\mathbf{z})})}{r(\mathbf{x},\mathbf{z}) + 1} = 1 \quad \text{if} \quad r(\mathbf{x},\mathbf{z}) \geq r(\mathbf{z},\mathbf{x}).$$

Therefore, with $\lambda(\mathbf{x},\mathbf{z}) = 1 + \min[r(\mathbf{x},\mathbf{z}), r(\mathbf{z},\mathbf{x})]$, we obtain $\alpha(\mathbf{x},\mathbf{z}) = \min[1, r(\mathbf{x},\mathbf{z})]$. Figure 2 provides a graphical representation of these connections among the two general forms Eq. (30) and (31) and their specific cases.

**Best choice.** Within this class of acceptance functions, the best choice following Peskun's ordering [79, 92] is the standard acceptance function $\min[1, r(\mathbf{x},\mathbf{z})]$. Indeed, recalling Eqs. (30)-(32) (and always we have $0 \leq \alpha \leq 1$), it is possible to show that

$$\alpha(\mathbf{x},\mathbf{z}) = r(\mathbf{x},\mathbf{z})\alpha(\mathbf{z},\mathbf{x}) \leq \min[1, r(\mathbf{x},\mathbf{z})], \tag{38}$$

i.e., roughly speaking, keeping fixed the proposal and target functions, the standard acceptance function provides higher acceptance rates.

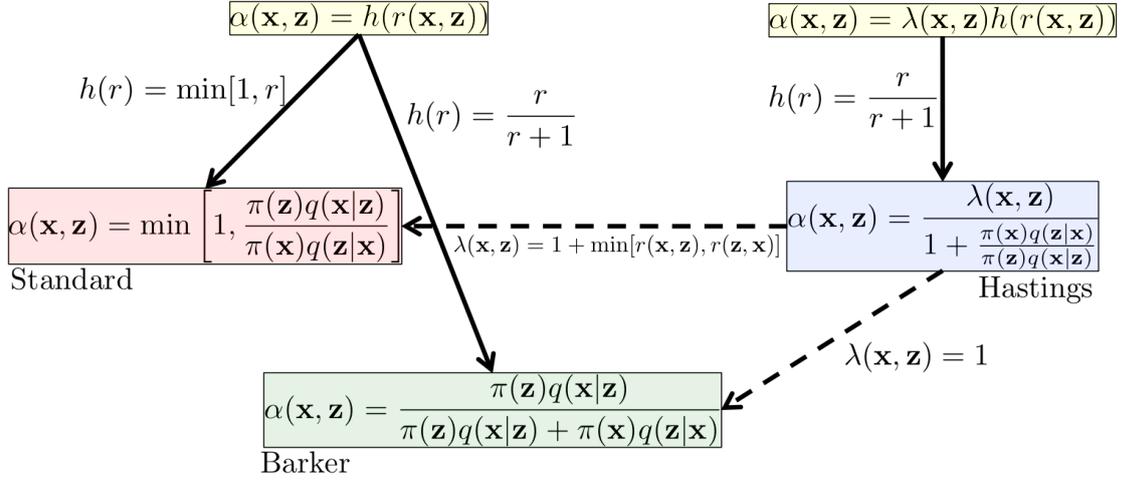

Figure 2: Relationships among the general forms of the acceptance function and the specific cases.

**Other examples.** Other possible alternatives have appeared. For instance, the following acceptance probabilities have been proposed in the literature,

$$\alpha(\mathbf{x}, \mathbf{z}) = \frac{\pi(\mathbf{z})}{q(\mathbf{z}|\mathbf{x})}\rho(\mathbf{x}, \mathbf{z}), \qquad \alpha(\mathbf{x}, \mathbf{z}) = \frac{1}{q(\mathbf{z}|\mathbf{x})\pi(\mathbf{x})}\rho(\mathbf{x}, \mathbf{z}), \qquad (39)$$

$$\alpha(\mathbf{x}, \mathbf{z}) = \frac{q(\mathbf{x}|\mathbf{z})}{\pi(\mathbf{x})}\rho(\mathbf{x}, \mathbf{z}), \qquad \alpha(\mathbf{x}, \mathbf{z}) = q(\mathbf{x}|\mathbf{z})\pi(\mathbf{z})\rho(\mathbf{x}, \mathbf{z}), \qquad (40)$$

where $\rho$ is a symmetric function, $\rho(\mathbf{x}, \mathbf{z}) = \rho(\mathbf{z}, \mathbf{x})$, such that $0 \leq \alpha(\mathbf{x}, \mathbf{z}) \leq 1$ for all $\mathbf{x}, \mathbf{z} \in \mathbb{R}^D$.

## 2.5 Acceptance rate

In the MH method, a tentative sample is drawn from a proposal distribution and then a test is carried out to determine whether the state of the chain should jump to the new proposed value or not. This test depends on the acceptance probability $\alpha$. If the jump is not accepted (with probability $1 - \alpha$), the chain remains in the same state as before. We can define the acceptance rate as

$$a_R = \int_{\mathcal{D}^2} \alpha(\mathbf{x}, \mathbf{z}) q(\mathbf{z}|\mathbf{x}) \bar{\pi}(\mathbf{x}) d\mathbf{z} d\mathbf{x} \approx \frac{1}{T} \sum_{t=1}^{T} \alpha(\mathbf{x}^{(t-1)}, \mathbf{z}^{(t)}), \qquad (41)$$

where the last expression is a Monte Carlo approximation of the integral in Eq. (41) where $\mathbf{x}^{(t-1)}$ represents the state of an MH chain at the $(t-1)$-th iteration, and $\mathbf{z}^{(t)}$ is the proposed sample at the $t$-th iteration,[4] i.e., $\mathbf{z}^{(t)} \sim \pi(\mathbf{z}|\mathbf{x}^{(t-1)})$. Clearly, $0 \leq a_R \leq 1$.

---

[4]Since the chain has $\bar{\pi}$ as invariant pdf, we have $\mathbf{x}^{(t-1)} \sim \bar{\pi}(\mathbf{x})$ after a "burn in" period. Namely, after a certain number of iterations, we have $(\mathbf{x}^{(t-1)}, \mathbf{z}^{(t)}) \sim q(\mathbf{z}^{(t)}|\mathbf{x}^{(t-1)})\bar{\pi}(\mathbf{x}^{(t-1)})$.

**Optimal acceptance rate $a_R^*$.** Given a target $\bar{\pi}(\mathbf{x})$ and choosing the class of the proposal functions to be used as $q_\sigma(\mathbf{x}|\mathbf{x}^{(t-1)})$, where $\sigma$ represents a scale parameter, there exists an optimal scale parameter $\sigma$ such that we obtain an optimal value $a_R^*$. This optimal acceptance rate $a_R^*$ minimizes the correlation among the samples within the chain. Unlike in a rejection sampler [36, 85], this optimal rate $a_R^*$ is in general unknown. Moreover, it varies depending on the specific problem, and in general differs from 1 (i.e., $a_R^* \neq 1$ in a generic sampling problem). In [86], under certain assumptions, the authors obtain $a_R^* \approx 0.234$ whereas, considering another scenario, the authors of [87, 88] obtain $a_R^* \approx 0.574$.

**Scenarios where $a_R \approx 1$.** Above, we have observed that the optimal acceptance rate $a_R^*$ generally differs from 1. However, in some scenarios, acceptance rates close to 1 could show that the proposal and the target functions have similar shapes and the MH sampler provides excellent performance. In other scenarios, it can mean catastrophic behaviors in terms of performance. Below, we list two different scenarios where we can have $a_R \approx 1$:

1. When the proposal coincides with the target density, i.e., $q(\mathbf{x}) = \bar{\pi}(\mathbf{x})$. This is clearly an ideal case, where the MH method is converted into an exact sampler, providing the optimal possible Monte Carlo performance, i.e., i.i.d. samples from the target $\bar{\pi}(\mathbf{x})$ (without considering negative correlation among the samples). In this case, $a_R = 1$.

2. When the scale parameter $\sigma$ of the proposal pdf $q_\sigma(\mathbf{x}|\mathbf{x}^{(t-1)})$ is very small w.r.t. the variance of the target. In this case the MH sampler tends to accept any proposed candidate in order to explore as quickly as possible the state space. The performance is often poor with high correlation among the generated samples.

In more sophisticated MH-type algorithms other similar scenarios exist. For instance, in certain advanced MCMC techniques, as the Multiple Try Metropolis (MTM) method [58] or the Independent Doubly Adaptive Metropolis Sampling (IA$^2$RMS) technique [69], if certain parameters grow to infinity, then $a_R \to 1$. In this case, the performance can be extremely good, providing virtually independent samples, but with an increased computational cost. The key point in these kind of methods is how they handle the trade-off between performance and computational cost. Another example where $a_R = 1$ occurs in an ideal (and only theoretical) scenario in the Hamiltonian Monte Carlo approach when the Hamiltonian equations can be solved analytically and no numerical integration steps (as the leapfrog method) are required [76].

## 3 Variants, Extensions and Enhancements

In this section, we provide a brief overview of the main variants, extensions and enhancements of MH which can be found in literature. We summarize the main ideas and classes of algorithms that have been widely used in different applications. This can be seen as a list of research topics which have been (or they still are) active research areas. Note that all the points in the following list generalize the elements and relax some of the assumptions described in Section 2.1 for the standard MH method. This section is summarized graphically in Figure 3.

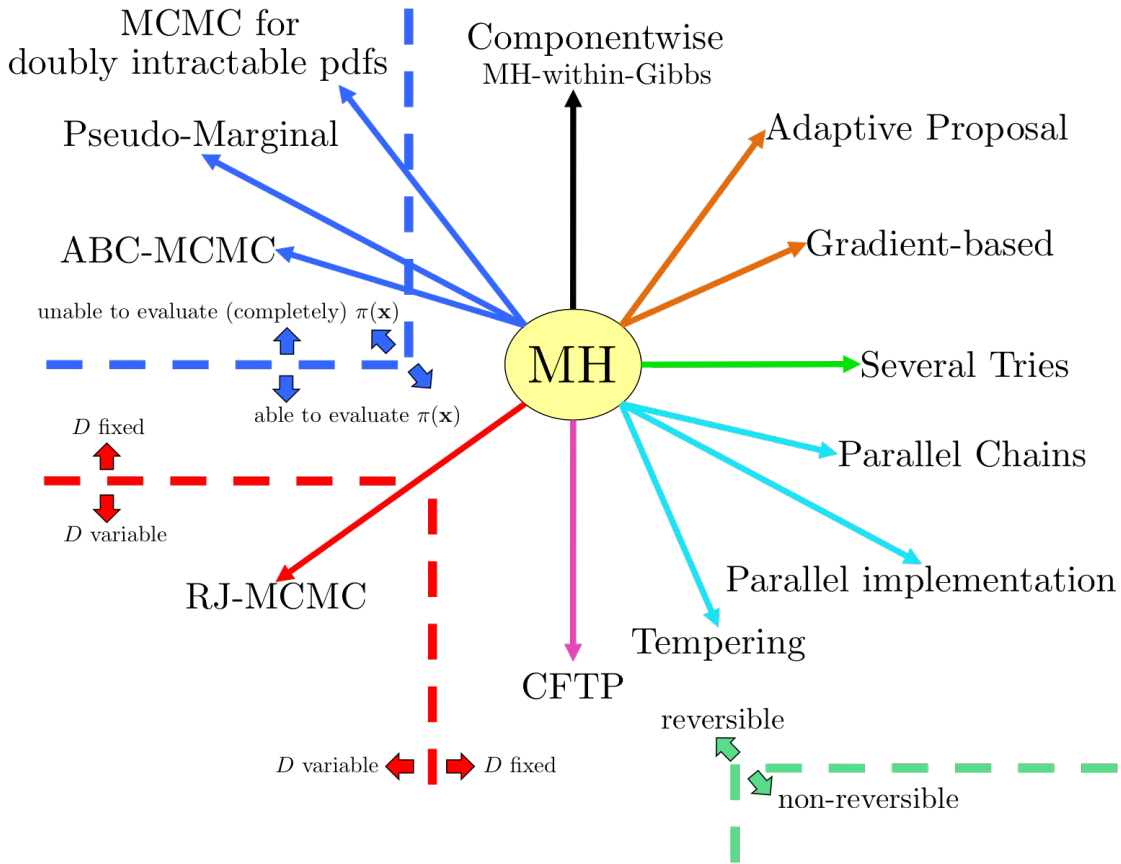

Figure 3: Graphical representation of different variants, extensions and research lines regarding the MH algorithm. In all these schemes, in general, **x** can be a discrete or continuous variable whereas the iteration $t$ of the chain is a discrete variable, typically $t \in \mathbb{N}$ (for continuous extensions see, e.g., [11]).

**Componentwise Metropolis sampling.** The standard MH technique works directly in the $D$-dimensional space of the variable of interest $\mathbf{x} = [x_1, \ldots, x_d, \ldots x_D]^\top \in \mathbb{R}^D$. However, componentwise approaches [41, 48, 54, 35] working iteratively in different uni-dimensional slices of the entire space (updating individually each component $x_d$) are possible. This scheme is called *Metropolis-within-Gibbs* or *Componentwise Metropolis* sampling [31, 56, 85]. In many applications, and for different reasons, the component-wise approach is the preferred choice. For instance, this is the case when the full-conditional distributions are directly provided or when the probability of accepting a new state with a complete block approach becomes negligible as the dimension of the problem $D$ increases. The Metropolis-within-Gibbs strategy can be applied with uni-dimensional slices (updating component by component) applying the MH for drawing from univariate full-conditional pdfs. Block-wise MH-within-Gibbs sampling approaches where several components are updated simultaneously have been also proposed.

**Improving the proposal and gradient-based approaches.** The performance of the MH technique depends strongly on the choice of the proposal density. With a proposal pdf closer to the target density, the algorithm provides a sequence of states with less correlation among each other and, as a consequence, Monte Carlo estimators with smaller Mean Square Error (MSE). Therefore, several schemes adapting the proposal function have been proposed [19, 45, 30, 23, 90]: some of them adapt parametric functions [40, 38, 60], whereas others consider non-parametric approaches [35, 69, 72]. The main issue with the adaptation of the proposal is to ensure the ergodicity of the resulting chain [43, 89].

Another way for improving the proposal is to incorporate any information available in advance about the target pdf. For instance, we can use the information provided by the gradient of the target posterior density: this is the idea in the *Metropolis-adjusted Langevin* algorithm (MALA) [87, 55, 85]. A more general approach, containing MALA as a special case, is given by the Hamiltonian Monte Carlo (HMC) schemes [76]. In HMC, the target pdf is extended by creating a complete artificial variable $\mathbf{p} \in \mathbb{R}^D$ (inspired by the momentum in physics) and the joint generalized target $\pi_g$ is defined following the Hamiltonian, i.e., $-\log \pi_g(\mathbf{x}) = -\log \pi(\mathbf{x}) + \frac{1}{2}\mathbf{p}^2$ [76] where $\frac{1}{2}\mathbf{p}^2$ plays the role of kinetic energy.

**Several tries per iteration.** In the MH technique, at each iteration one new sample $\mathbf{z}'$ is generated to be tested with the previous state $\mathbf{x}^{(t-1)}$ by the acceptance probability $\alpha(\mathbf{x}^{(t-1)}, \mathbf{z}')$. Other generalized MH schemes generate several candidates at each iteration to be tested as new possible state. In all these schemes, the acceptance probability $\alpha$ is properly designed in order to guarantee the ergodicity of the chain. The *Multiple Try Metropolis* (MTM) algorithms are examples of this class of methods [58, 21, 22, 68], where $N$ samples are drawn from the proposal pdf, then one of them is selected according to some suitable weights, and finally the selected candidate is accepted as new state according to a generalized probability function $\alpha$. A similar approach is the so-called *Ensemble MCMC* (EnMCMC) technique [75, 20, 65]. When an independent proposal $q(\mathbf{z})$ is employed, the EnMCMC scheme can be seen as an MH generalization using an extended Barker acceptance function (see Section 2.4). Moreover, both MTM and EnMCMC can be also interpreted as a way to combine an MCMC approach with the resampling procedure [68, 65]. Another interesting scheme (possibly) drawing different candidates per iteration is the Delaying Rejection MH (DRMH) algorithm [94]. In DRMH, if the generated sample $\mathbf{z}'_1$ is rejected according to the probability $\alpha_1$, then a new candidate $\mathbf{z}'_2$ is drawn and tested according to a new suitable probability $\alpha_2$ (ensuring the ergodicity). If $\mathbf{z}'_2$ is rejected the process continues, generating a new sample $\mathbf{z}'_3$. The information provided by the rejected tries $\mathbf{z}'_1, \mathbf{z}'_2$ can be used for improving the proposal function. If $\mathbf{z}'_2$ is accepted, then the chain is moved forward. The main advantage of all these schemes (MTM, EnMCMC and DRMH) is that they can explore a larger portion of the sample space, at the expense of a higher computational cost per iteration (more evaluations of the target are needed at each iteration).

The *particle Metropolis-Hastings* (PMH) method [2] belongs to the MTM family: in this case, the $N$ samples (called *particles*) and the corresponding weights are generated sequentially via sequential importance resampling, a.k.a., particle filtering technique [62, 66]. The presence of the resampling steps adds correlation among the $N$ tries (particles) but the ergodicity is not jeopardized (note also that several MTM schemes with correlated tries have been proposed in literature [22, 67]). In the so-called *particle marginal MH* (PM-MH) algorithm [2], a subset of parameters is generated sequentially and the rest of the parameters are drawn with a standard block approach. This fits particularly well for making inference in state space models where the dynamical variables are drawn by a particle filter and the unknown (static) parameters are generated in a standard way (using a proposal pdf with a block approach). More specifically, the PM-MH performs as an MTM scheme w.r.t. to the dynamic parameters, whereas PM-MH performs as a standard MH w.r.t. the static parameters [66]. Furthermore, PMH and MTM using an independent proposal pdf can be also interpreted as standard MH schemes when an extended importance sampling theory is employed [62].

**Parallel chains, tempering of the target, and parallel implementation.** For a fixed computational cost, the use of a longer single MH chain should be better than the use of several shorter independent runs, in terms of convergence (in order to exceed the burn-in period). However, in certain scenarios (as a multimodal target pdfs, for instance), there is not a clear advantage of using single longer chain with respect to employing independent parallel chains (IPCs) [50, 65]. IPCs can foster the exploration of the state space and improve the overall performance. Several studies suggest different ways for exchanging information among the parallel chains that, in this case, are no longer independent [65, 23, 34]. These schemes are often combined with the idea of defining an extended target

$$\bar{\pi}_g(\mathbf{x}_1, \ldots, \mathbf{x}_R) \propto \prod_{r=1}^{R} \pi_r(\mathbf{x}_r), \quad (42)$$

where $\pi_1(\mathbf{x}_1) = \pi(\mathbf{x})$ is the original target and the other ones are tempered (i.e., scaled) versions of $\pi$, i.e., $\pi_r(\mathbf{x}_r) = [\pi(\mathbf{x})]^{\gamma_r}$ (unlike for optimization, in this case $0 < \gamma_r < 1$ for all $r$).[5] In this type of scheme, $R$ different chains are performed considering a different target function $\pi_r(\mathbf{x}_r)$, and they exchange information in order to improve the mixing of the first one addressing $\pi(\mathbf{x})$. Another related topic is the parallel implementation of the standard MH algorithm, distributed across different processors/machines. The MH method has traditionally been implemented in an iterative non-parallel fashion. Several authors have studied its parallelization, in order to reduce their computation time [15, 46].

**Variable dimension $D$ of the inference space (inferring $D$).** Several applications involve varying-dimension problem. Hence, it is necessary that the space of the parameters can change its dimension $D$ ($\mathbf{x} \in \mathbb{R}^D$). For instance, this usually occurs for the joint purpose of parameter estimation and model selection, e.g., when the number of parameters in the model is unknown. Typically, more complex models require more parameters whereas simpler models require fewer parameters. We can interpret that the dimension $D$ is an additional parameter that has to be inferred. The *Reversible Jump MCMC* (RJ-MCMC) algorithm proposed in [39] is an extension of the MH method capable of jumping between spaces of different dimensionality. This is possible with a suitable generalization of the acceptance probability function $\alpha$ (incorporating the Jacobian of the transformation considered for connecting the different spaces).

---

[5]The dimension of the problem is increased, but it is considered fixed during the inference.

**When the posterior cannot be evaluated.** In some applications, it is not possible to evaluate the target function $\pi(\mathbf{x})$. In some cases, we can compute an extended target $\pi_e(\mathbf{x}, \mathbf{v})$ having $\pi(\mathbf{x})$ as marginal pdf, and we are not interested in inferring the parameter $\mathbf{v}$ but only $\mathbf{x}$. If we cannot evaluate directly the marginal $\pi(\mathbf{x})$, we have to approximate this marginal pdf in some way. The approximated MH methods for this scenario are called *pseudo-marginal algorithms* [8, 3]. A related problem (as noted, e.g, in [82]) appears when the normalizing constant of the likelihood function (a.k.a., partition function) is unknown and depends to the parameter $\mathbf{x}$ to be inferred. Namely, a piece/factor of the likelihood function cannot be evaluated and must be estimated as well. This is the case of the *doubly-intractable distributions* [82] and the *Møller method* [73] and the so-called *exchange algorithm* [74]. When the entire likelihood $\ell(\mathbf{y}|\mathbf{x})$ is completely unknown or cannot be evaluated, the so-called *Approximate Bayesian Computation* (ABC) approach can be applied [9, 61]. In both cases, it is necessary to generate artificial data $\mathbf{y}_1, \mathbf{y}_2, ..., \mathbf{y}_M \sim \ell(\mathbf{y}|\mathbf{x}')$ given a generic parameter value $\mathbf{x}'$. This task can require the use of another Monte Carlo technique (for instance, a rejection sampler or other MCMC algorithm).

**Convergence detection.** Several studies have been devoted to the detection of the convergence or lack thereof (establishing convergence bounds etc.) [29, 85]. One algorithm that avoids this problem is the *Coupling From The Past* (CFTP) technique [81]. However, its design and implementation tend to require a great care. In general, implementing CFTP for practical application is still a difficult task.

**Non-reversible algorithms.** The standard MH method, as well as several its generalizations, satisfies the detailed balance condition so that the produced chain is reversible (see Section 2.3). However, the reversibility is only used as a theoretical tool in order to show that the generated chains converges to the target distribution (recall that is a sufficient condition but not necessary). However, different studies in the literature show that non-reversible Markov chains may have better properties in terms of mixing behavior or asymptotic variance, hence providing better performance [12, 77].

# Acknowledgements

This work has been supported by the European Research Council (ERC) through the ERC Consolidator Grant SEDAL ERC-2014-CoG 647423.